\documentclass[aps,prl,reprint,superscriptaddress]{revtex4-1}

\usepackage{amsmath,amssymb}
\usepackage{graphicx}% Include figure files
\usepackage{bm}% bold math
\usepackage[latin1]{inputenc}
\usepackage[caption=false]{subfig}
\RequirePackage{hyperref}
\usepackage{color}
\usepackage{soul}

\begin{document}

\title{Dynamical trapping through harmonic breathing}

\author{Sebasti\'an Carrasco}
\email[]{sebastian.carrasco@ug.uchile.cl}
\affiliation{Departamento de F\'isica, Facultad de Ciencias,
  Universidad de Chile, Casilla 653, Santiago, Chile 7800024} 

\author{Jos\'e Rogan}
\affiliation{Departamento de F\'isica, Facultad de Ciencias,
  Universidad de Chile, Casilla 653, Santiago, Chile 7800024} 
\affiliation{Centro para la Nanociencia y la Nanotecnolg\'ia, CEDENNA,
  Chile}

\author{Juan Alejandro Valdivia}
\affiliation{Departamento de F\'isica, Facultad de Ciencias,
  Universidad de Chile, Casilla 653, Santiago, Chile 7800024} 
\affiliation{Centro para la Nanociencia y la Nanotecnolg\'ia, CEDENNA,
  Chile}

\date{\today}

\begin{abstract}
  \noindent A new strategy for trapping quantum particles is
  presented, which behaves like an effective harmonic oscillator
  potential trap wherever is desired. The approach is based on
  harmonic contraction and expansion of the system around a fixed
  point (trapping point) at high frequencies.  Analytical results are
  presented for an arbitrary potential and contrasted with numerical
  calculations for a quantum particle between two impenetrable
  walls. Similarly, making use of the analogy between quantum
  mechanics and optics, it is shown that an harmonically breathing
  waveguide lattice can be used to trap optical beams using
  frequencies of the order of the coupling constant without needing to
  resort to nonlinear dielectric terms.
\end{abstract}

\pacs{36.40.-c, 36.40.Ei, 36.40.Qv, 36.40.Mr, 36.40.Sx, 61.46.-w,
  63.22.Kn, 82.30.Nr.}

\maketitle

Reaching and stabilizing novel quantum states using controllable
quantum systems is a common goal of many fields of physics. For
example, in the last few years, there has been some interesting
suggestions in the literature, based on geometric or topological
restrictions, on how to accomplish such a
goal~\cite{rechtsman2013photonic, polini2013artificial,
  jacqmin2014direct}. Similarly, there has been other approaches that
are based on driving a system using external fields or mechanical
deformations to dynamically generate such desired properties
\cite{dalibard2011colloquium, zhai2012spin}. This strategies allows
the exploration of situations that go beyond the static capabilities.
In that context, the alliance produced by the correspondence between
quantum mechanics and optics has given rise to a resourceful
laboratory where such driven systems can be experimentally tested, in
what can be called ``quantum simulation''. Examples include dynamic
localization \cite{longhi2006observation, szameit2010observation,
  longhi2015localization}, coherent destruction of tunneling
\cite{longhi2005coherent, della2007visualization,
  szameit2009inhibition, mukherjee2016observation, luo2007nonlinear},
Rabi oscillations \cite{ornigotti2008visualization,
  PhysRevLett.99.233903, PhysRevLett.102.123905}, Anderson
localization \cite{schwartz2007t, wiersma1997localization,
  lahini2008anderson, wolf1985weak, van1985observation,
  john1984electromagnetic}, and dynamical trapping
\cite{theocharis2006dynamical, longhi2011dynamic,
  papagiannis2011power}, among others.

Many of such examples have been optically simulated by exploiting the
equivalence between a sinusoidally curved waveguide and an AC field
\cite{longhi2005self, longhi2006observation, iyer2007exact} at the
quantum level. Other examples, specifically in the context of dynamic
stabilization, was presented by Stefano Longhi in
Ref. \cite{longhi2011dynamic}, which was based on a periodic
graded-index modulation of a wave-guided lattice in the longitudinal
direction. His approach works in a similar way to the Kapitza (or
dynamic) stabilization effect of classical and quantum mechanics in
rapidly oscillating potentials; or Paul traps for charged particles
that appeals to rapidly oscillating potential to trap a particle in
cases where the static potential cannot \cite{earnshaw1842nature}.

In this manuscript, we present a new strategy to make dynamic
stabilization (or trapping) using a harmonically breathing potential,
in other words, harmonically expanding and contracting the potential
around a fixed point, which plays the role of the trapping point, all
of this in analogy with the Kapitza stabilization stabilization in
classical mechanics.  Moreover, here is shown analytically for an
arbitrary potential that such breathing system can behave exactly like
an effective harmonic oscillator potential in the high-frequency
limit, and the trapping region can be directly controlled. We offer
two examples as an illustration of the trapping effect.  First, we
consider a quantum system where the potential is given by two
impenetrable walls, and second, an optic waveguide lattice. In both
systems, the dynamic stabilization is present allowing to trap the
quantum particle and the light beam where is desired, but the
simplicity of the first system allows us to consider numerically the
effective harmonic oscillator potential trap limit. For the waveguide,
is shown that the trapping effect persists even for relatively small
frequencies and amplitudes, making this effect suitable for
experimental examination.

Let us consider a quantum particle under the influence of a breathing
potential. The evolution of the wavefunction follows the Schr\"odinger
equation, namely
\begin{equation} \label{wave}
  i \hbar \frac{\partial \psi}{\partial t} = - \frac{\hbar^2}{2 m} \frac{\partial^2 \psi}{\partial x^2} + \frac{1}{\alpha^2} V \left(\frac{x}{\alpha}\right) \psi \ ,
\end{equation}
where $t$ is the time, $x$ the spatial position, $m$ the mass, and
$\alpha=\alpha(z)$ describes the breathing of the potential $V$. By
means of the transformation to the non-breathing frame, namely $x' = x
/ \alpha(z)$, $t' = t$, and $\phi(x', t') = \psi(x', z') \, e^{- i
  \alpha \dot \alpha x'^2 / (2 \hbar)}$ (where the dot indicates the
derivative with respect to $t'$) the previous equation \eqref{wave}
reads
\begin{equation} \label{wave2}
  i \hbar \frac{\partial \phi}{\partial t'} = - \frac{\hbar^2}{2 m \alpha^2} \frac{\partial^2 \phi}{\partial x'^2} + \frac{V(x')}{\alpha^2} \phi -\left(\frac{1}{2} \alpha \ddot \alpha + \dot \alpha^2 \right) m x'^2 \phi \ .
\end{equation}
So far the treatment is exact, but let's suppose that $\alpha = 1 +
\epsilon \cos (\omega z')$, where $\epsilon$ is an adimensional
parameter which measures the breathing amplitude and $\omega =
\frac{2\pi}{T}$ is the breathing frequency and $T$ the breathing
period. Note that $|\epsilon| < 1$ to avoid singularities. For large
values of $\omega$ and small values of $\epsilon$ Eq. \eqref{wave2}
becomes
\begin{equation} \label{wave3}
  i \hbar \frac{\partial \phi}{\partial t'} = -
  \frac{\hbar^2}{2 m} \frac{\partial^2 \phi}{\partial x'^2} +
  V(x') \phi + \frac{1}{2} m \epsilon \omega^2 \cos (\omega t') x'^2
  \phi \ .
\end{equation}
Which is a harmonic trap, and so the quantum particle is expected to
be contained close to $x' = 0$. Moreover, the evolution in this limit
(see ref. \cite{goldman2014periodically}) can be approximated by
\begin{equation} \label{Evolution}
  i \hbar \frac{\partial \phi}{\partial t'} = - \frac{\hbar^2}{2 m} \frac{\partial^2 \phi}{\partial x'^2} + V(x') \phi + \frac{1}{2} m \Omega^2 x'^2 \phi \ ,
\end{equation}
with $\Omega = \epsilon \omega / \sqrt{2}$. Similarly, in this limit
the term proportional to $\dot{\alpha}^2$ should give a much smaller
contribution. Note that the trapping could be done wherever is
desired, namely $x_0$, using $x' = (x - x_0)/ \alpha$ instead of $x' =
x/\alpha$. Furthermore, this result is very promising, because in many
other strategies the strength of the confining part of the effective
potential usually scales as $\sim 1/\omega^2$ \cite{cook1985quantum,
  gilary2003trapping}, and so the observation of the stabilizing
dynamics might require extremely long propagation distances. However,
in our approach, it grows with $\omega^2$ reducing the
required propagation distance to observe the trapping, as
we will discuss further.
\begin{figure}
  \includegraphics[scale=.7]{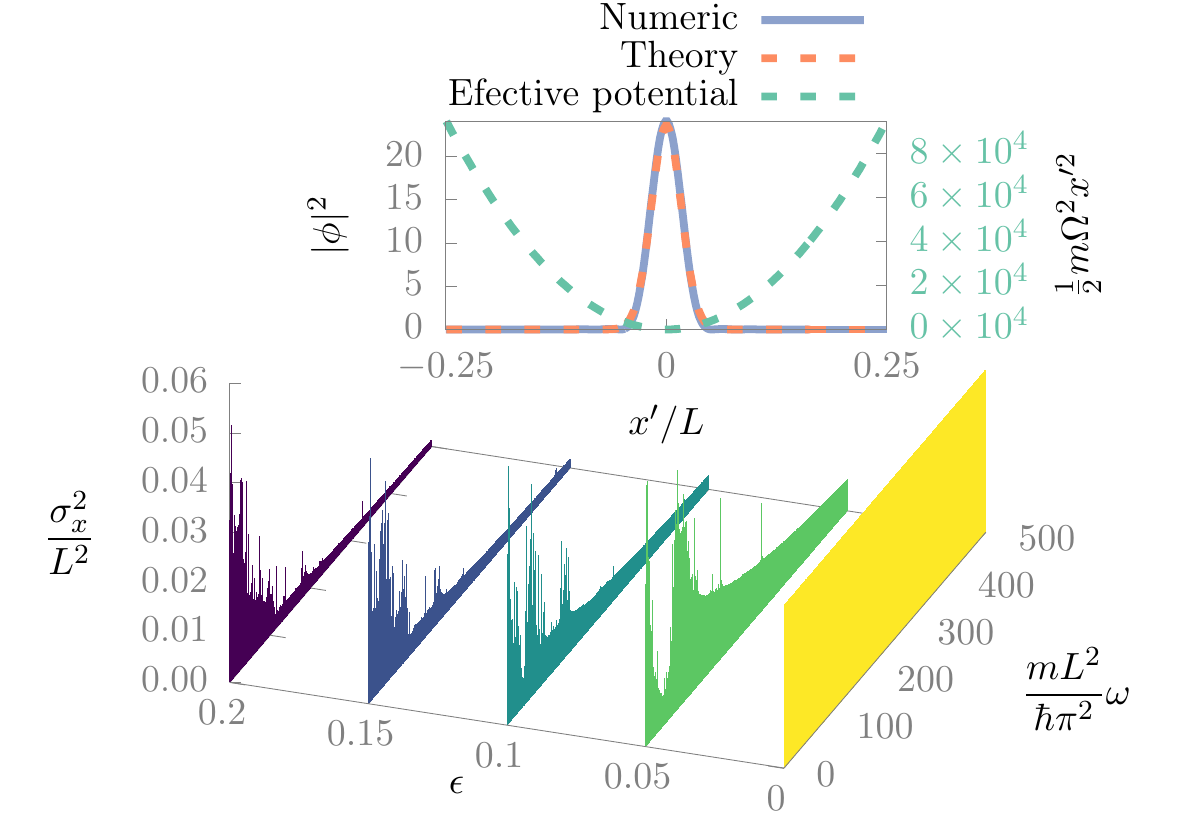}
  \caption{\label{average} Dynamical trapping produced by the
    breathing of \label{average} two impenetrable walls. (a) The
    numerically computed variance of the Floquet eigenstate with the
    lowest variance as a function of $\omega$ and $\epsilon$. (b)
    Comparison between the lowest variance Floquet eigenstate and the
    ground state predicted by the average harmonic oscillator
    potential trap, and the corresponding effective potential produced
    by the breathing of the system for $\omega = 5,000 \, \hbar \pi^2
    m^{-1} L^{-2}$, and $\epsilon = 0.05$. The numerical calculation
    was performed up to $N = 30$ eigenstates of the static problem.}
\end{figure}

First, and for the sake of illustration (both numerical and
analytical) let's show the stabilization effect and the validity of
our approximation assuming that the potential is the one given by two
impenetrable walls at $x = -L/2$ and $x = L/2$, and we proceed to
solve the lowest variance Floquet eigenstate. The Floquet eigenstates
$\phi^{(\mu)}$ and the quasi-energies or Floquet exponents $\mu$ are
defined as the solutions of Eq. \eqref{Evolution} of the form
$\phi^{(\mu)}(z'+T) = \phi^{(\mu)}(z') \, \exp(-i \mu T)$. uch states
are calculated expanding and projecting $\phi$ in the eigenfunctions
$\phi_n$ of the static case ($\epsilon = 0$), namely, $\phi =
\alpha^{-1/2} \sum_{n = 1}^{\infty} a_n \phi_n$, and truncating the
sum up to $N$ to obtain system of ordinary differential equations. The
factor $\alpha^{-1/2}$ should be added in order to preserve
normalization.  In Fig. \ref{average} we observe the dynamical
trapping produced by the breathing waveguide.  This is revealed by
Fig.  \hyperref[average]{\ref*{average}(a)}, where we see the
numerically calculated values of $\sigma_x^2$, the variance in the
coordinate $x$ of the light envelope, for the Floquet eigenstate, with
the lowest variance, as a function of $\epsilon$ and $\omega$. Note
that generally, the envelope becomes more localized as $\epsilon$ and
$\omega$ grows, as it has been predicted by the
Eq. \eqref{Evolution}. However, we can observe some exceptions as
peaks of $\sigma_x^2$ for certain spatial frequencies.  These peaks
are associated with resonant behavior between the normal modes of the
static system. In such case, a lowest variance Floquet eigenstate is
expected to be a linear combination of some of such modes that include
the resonant one, and so is likely to be an unlocalized
state. However, such effects are more clearly observed at low
frequencies because the resonance region becomes much thinner as the
difference of frequencies becomes larger (see Ref.
\cite{carrasco2017}). Even more, we can conclude that it is more
likely to encounter the system localized because such exceptions only
occur at very specific frequencies, and so the effect is very
robust. In Fig.  \hyperref[average]{\ref*{average}(b)} we show the
lowest variance Floquet eigenstate of the full Eq.~\eqref{wave2}, at
$\omega = 25 \, \hbar \pi^2 m^{-1} L^{-2}$ and $\epsilon = 0.05$, and
the one predicted by Eq.  \eqref{Evolution}, and the corresponding
effective harmonic oscillator potential trap produced by the breathing
of the system.  As we can see, our approximation agrees quite closely
with the exact solution, even for not too large spatial frequencies
(tens of times the frequency of the lowest mode of the non breathing
impenetrable walls). Summing up, we found the predicted trapping
effect, which is still present at relatively low spatial frequencies,
but may be destroyed by resonances with the Rabi oscillations between
the static modes.

\begin{figure}
  \centering \includegraphics[scale=.7]{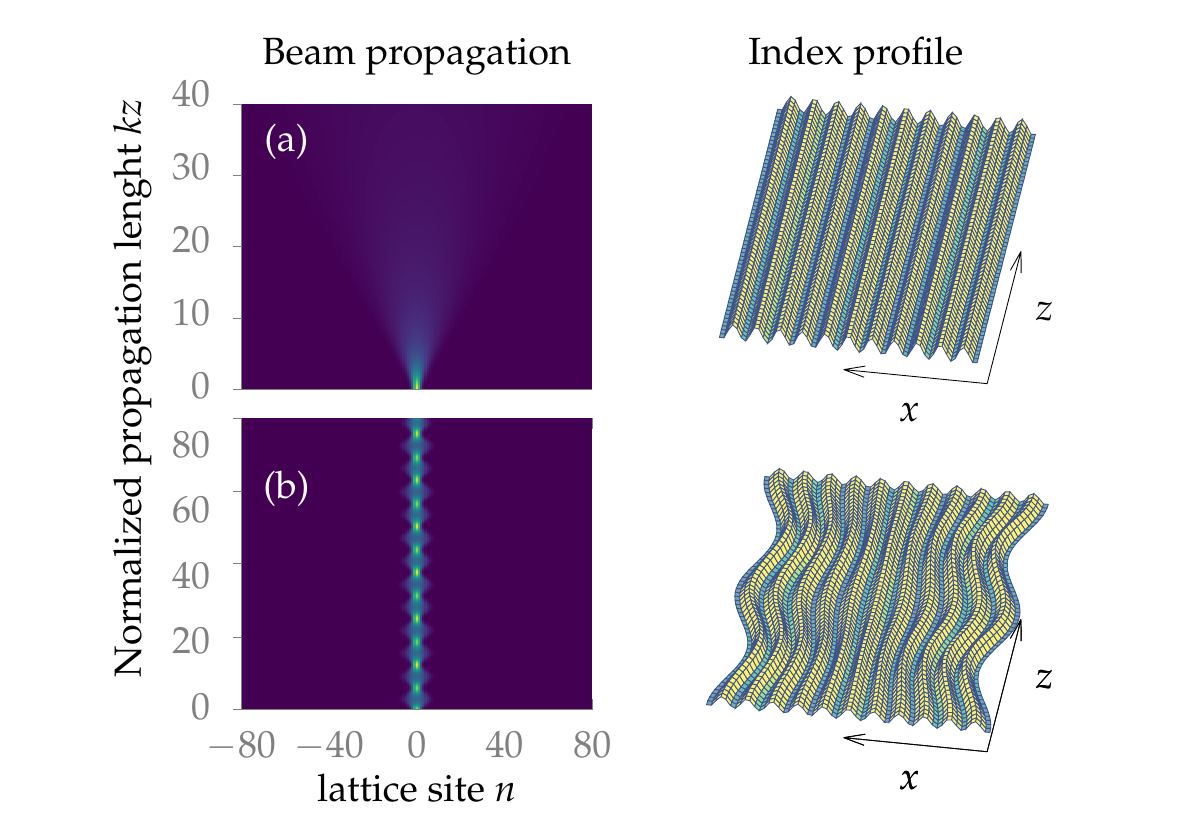}
  \caption{\label{beam} Beam propagation (snapshot of
    $|c_n(z)|^2$) in a breathing lattice with $\omega = k$ for (a)
    $\epsilon = 0$ and (b) $\epsilon = 0.1$. The right panels
    schematically show the index profile of the non-breathing (upper
    plot) and breathing (lower plot) waveguide array.}
\end{figure}

Now we turn to another important objective of this work, namely, to
show how breathing lattice can trap light beams where non-breathing
lattices cannot. Let us consider light propagating in a waveguide
which is breathing in the perpendicular direction $x$, and propagates
along the parallel direction $z$. The time evolution of a complex beam
envelope function $\psi(x, z)$ follows an optical analog of the
Schr\"odinger equation, and so following the same arguments presented
above, the optical equivalent of the Eq. \eqref{wave2}
with $m$ exchanged for $n_s$, $t'$ for $z'$, and $\hbar$ for
$\lambdabar$ is
\begin{equation} \label{wave2_op}
  i \lambdabar \frac{\partial \phi}{\partial z'} = - \frac{\lambdabar^2}{2 n_s \alpha^2} \frac{\partial^2 \phi}{\partial x'^2} + \frac{V(x')}{\alpha^2} \phi -\left(\frac{1}{2} \alpha \ddot \alpha + \dot \alpha^2 \right) n_s x'^2 \phi \ ,
\end{equation}
where $z'$ is the paraxial propagation distance, $x'$ now is the
spatial position in the non-breathing frame, $\lambdabar = \lambda /
(2 \pi)$ is the reduced wavelength, $V(x) = n_s - n(x)$, $n(x)$ is the
effective refractive index profile of the array, $n_s$ is the bulk
refractive index, and $\alpha=\alpha(z)$ describes the breathing of
the waveguide. Now let us consider a breathing waveguide lattice,
which in the standard nearest-neighbor tight-binding approximation is
governed by a discrete version of Eq. \eqref{Evolution} (see
Appendix), namely
\begin{equation} \label{discrete}
  i \dot c_n = - \frac{k}{\alpha} ( c_{n+1} + c_{n-1} ) + i \frac{\dot \alpha}{\alpha} c_n - \left(\frac{1}{2} \alpha \ddot \alpha + \dot \alpha^2 \right) \frac{n_s a^2 n^2}{\lambdabar}  c_n \ ,
\end{equation}
where $k$ is the coupling constant between adjacent waveguides, $a$
the distance between adjacent waveguides in the non-breathing case,
and $c_n$ the mode amplitude. Fig.  \ref{beam} shows the propagation
of a Gaussian beam obtained by direct numerical simulations of
Eq. \eqref{discrete} with $\alpha(z) = 1 + \epsilon \cos(\omega z)$,
as it was chosen for the previous calculations. It worth to recall
that the no singularity condition ($|\epsilon| < 1$) keeps the
distances between two points always greater than zero, hence two
waveguides never touch each other. The initial condition is set as
$c_n(0) = \exp(-n^2/5)$. In Fig. \hyperref[average]{\ref*{beam}(a)} we
observe the propagation through a non-breathing ($\epsilon = 0$)
waveguide lattice which acts as a defocusing lens for the discretized
beam. In Fig. \hyperref[average]{\ref*{beam}(b)} we observe the
dynamic trapping of the beam as it travels through the breathing
waveguide with $\epsilon = 0.1$, $\omega = k$, and $n_s a^2
\lambdabar^{-1} k = 1$, which is very close of to the experimental
case, considering that the typical experimental values, namely $a$ of
the order of $10 \, \mu\text{m}$, $\lambdabar$ of the order of $100 \,
\text{nm}$, $n_s$ around unity, and $k$ of the order of
$\text{mm}^{-1}$ \cite{longhi2005self, longhi2006observation,
  weimann2018decay}.  Indeed, for this typical coupling constant $k$
is of the order of a $\text{mm}^{-1}$, and the propagation length
shown on Fig. \ref{beam} corresponds to a physical length which is of
the order of 1 $\text{cm}$.  Remarkably, the effect which was
illustrated for the impenetrable wall problem at high frequencies
persists even for relatively small spatial modulation frequencies (of
the order of the coupling constant $k$), opening the possibility to
observe this trapping effect with the current experimental
capabilities.

\begin{figure}
  \centering
  \includegraphics[scale=.7]{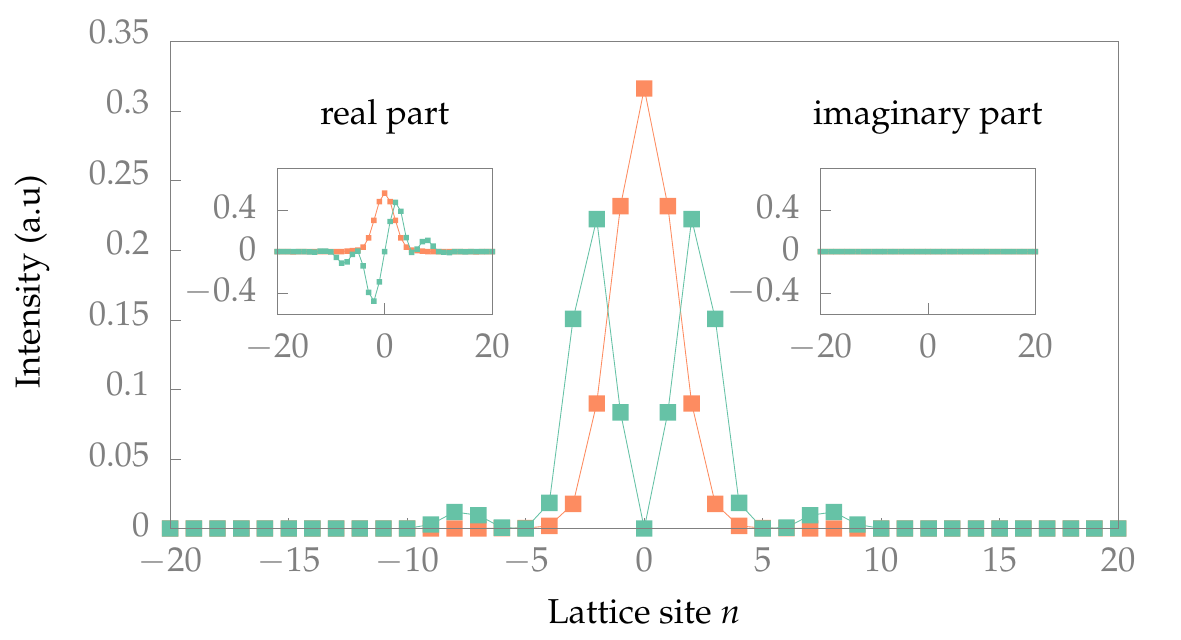}
  \caption{\label{discreto_floquet} Numerically computed intensity
    distribution $|c_n^2|$ for the two most localized metastable
    states (Floquet eigenstates), at the plane $z = 0$, for the
    breathing waveguide lattice with $\omega = k$ and $\epsilon =
    0.1$. The insets show the real and imaginary parts of the
    states.}
\end{figure}
This trapping effect is related to the existence of a Floquet
eigenstate (metastable state) which is similar to the initial
condition, or a sort of an effective potential produced by the
breathing of the system, as it was discussed before for an arbitrary
potential. For the parameters of the breathing waveguide used in Fig.
\hyperref[average]{\ref*{beam}(b)}, the Floquet eigenstates can be
numerically calculated. Assuming an array of 161 waveguides, in
Fig. \ref{discreto_floquet} we observe the two most localized
states. It's interesting to note that the most localized metastable
state highly resembles the initial condition used in
Fig. \hyperref[average]{\ref*{beam}(b)}, and then explain why it
remains trapped around the lattice site $n =0$ while it travels along
the propagation axis $z$.

\begin{figure}
    \vspace{- 1.8 cm}
  \centering
  \includegraphics[scale=.7]{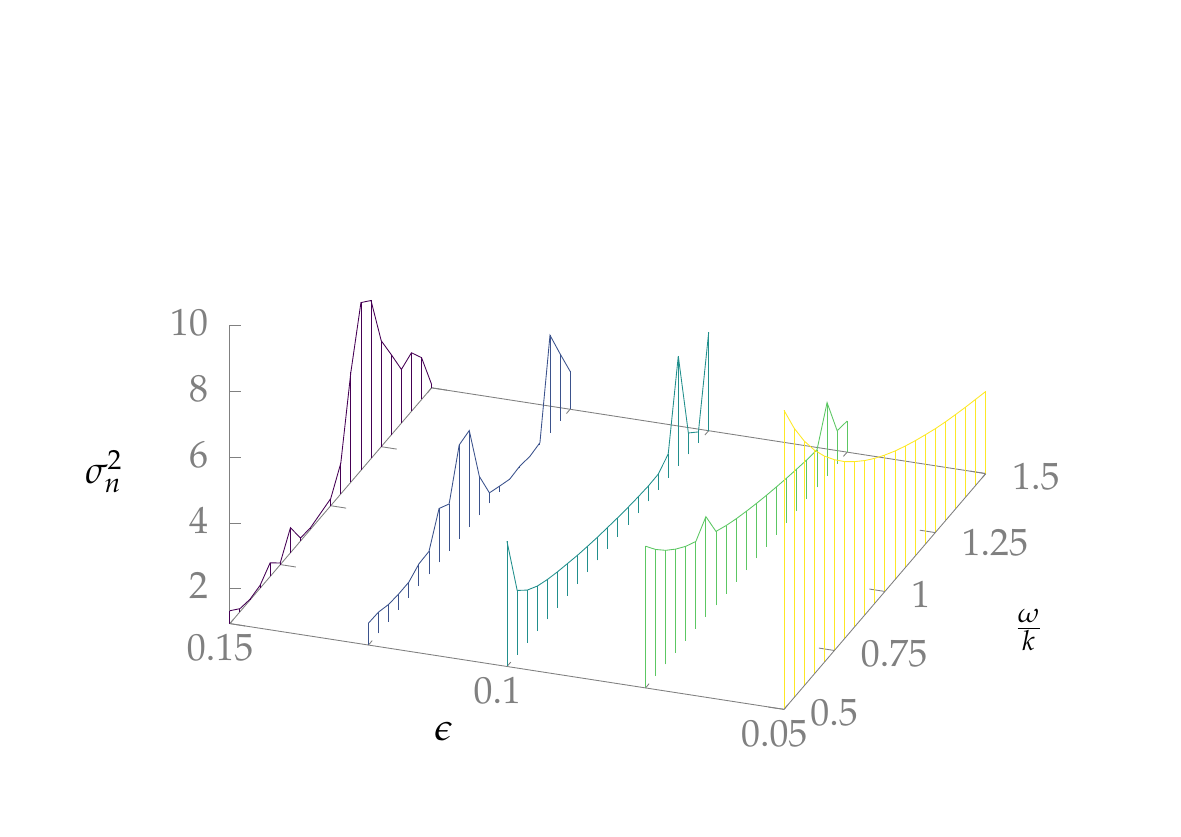}
  \caption{\label{desv_discreta} Numerically computed lowest variance
    of all Floquet eigenstates in the waveguide lattice as a function
    of $\omega$ and $\epsilon$.}
\end{figure}

In a similar way that it was done before for the case of two
impenetrable walls, in Fig. \ref{desv_discreta} we see the numerically
calculated values of $\sigma_n^2 = \sum_{n}^{} n^2 c_n^2$, the
variance in the lattice site $n$ of the mode amplitude, and plotted
the lowest variance among all Floquet eigenstates as a function of
$\epsilon$ and $\omega$. We observe that the behavior of the waveguide
lattice strongly resembles the behavior of the system of two
impenetrable walls, and so the localization grows as we increase the
value of the breathing amplitude $\epsilon$ and the spatial frequency
$\omega$. However, this doesn't hold for certain frequencies and
values of $\epsilon$ which, as it was stated before, may correspond to
a Rabi-like behavior of the system. As a final remark, we must state
that this show again that the effect persist even for relatively small
spatial modulation frequencies.

In conclusion, a new scheme of dynamical stabilization is presented,
in a way that is directly based in the simulation of a Paul trap
\cite{goldman2014periodically} using a harmonically breathing
potential at high frequency. Using this approach it's possible to trap
light wherever is desired, and such capability remains even for
relatively small breathing frequencies. Even more, such trapping
allows us to simulate potentials that may go beyond the static
experimental capabilities. Finally, it must be stated that this
particle trapping strategy analogy is not limited to quantum particles
and light, it can be applied to any system that follows a
Schr\"odinger-like equation, and so it can be used equally to trap
other kinds of bosons and fermions.

The authors are pleased to acknowledge M. Clerc for a valuable
discussion about stabilization effects. This work was supported by
CONICYT-PCHA/Doctorado Nacional/2016-2116103 (S.C.) and the Fondo
Nacional de Investigaciones Cient{\'\i}ficas y Tecnol\'ogicas
(FONDECYT, Chile) under grants \#1160639 (JR), \#1150718 (JAV), and
CEDENNA through ``Financiamiento Basal para Centros Cient{\'\i}ficos y
Tecnol\'ogicos de Excelencia-FB0807'' (JR and JAV). We also
acknowledge support from Grant FA9550-18-1-0438 of the U.S.A.F. Office
of Scientific Research (JR and JAV).

\emph{Appendix.---}In the tight-binding approximation
\begin{equation} \label{phi}
  \phi = \sum_{n} c_n u_n \left(x\right) \ ,
\end{equation}
where $u_n(x) = u(x-na)$ is a localized mode in the lattice site
$n$. It is reazonable to assume that the localized mode is Gaussian,
namely
\begin{equation*} \label{}
  u(x) = \frac{1}{(2 \pi b^2)^{1/4}} \exp\left(- \frac{x^2}{4 b^2}\right)  \ .
\end{equation*}
In the nearest-neighbor approximation
\begin{equation*} \label{}
  \int H_0 u_n u_{m} \, dx \approx e \delta_{n,m} - \lambdabar k \delta_{n, m+1} \ ,
\end{equation*}
where $H_0$ is the Hamiltonian in the non-breathing case, $e$ the
energy of the localized mode, and $k > 0$ is the coupling constant
between adjacent waveguides. It will be usefull to use the fact that
in the nearest-neighbor approximation
\begin{equation*} \label{}
  \int x^2 u_n u_m \, dx \approx n^2 a^2 \delta_{n, m} \ .
\end{equation*}
Now let's turn to the breathing case. The time evolution follows Eq \eqref{wave2_op}, namely
\begin{equation} \label{timev}
  i \lambdabar \frac{\partial \phi}{\partial z'} = \frac{1}{\alpha} H_0 \phi - \left(\frac{1}{2} \alpha \ddot \alpha + \dot \alpha^2\right) n_s x'^2 \phi \ .
\end{equation}
In the non-breathing frame Eq. \eqref{phi} should be rewritten as
\begin{equation*} \label{}
  \phi = \sum_{n} \sqrt{\frac{1}{\alpha}} c_n u_n \left(x'\right) \ .
\end{equation*}
Note that a factor $\sqrt{1/\alpha}$ should be added to preserve the
normalization. Reeplacing this back on \eqref{timev} and projecting,
gives
\begin{multline*} \label{}
  i \dot c_n = - \frac{k}{\alpha} (c_{n+1} + c_{n-1}) + i \frac{\dot \alpha}{\alpha} c_n + \frac{e}{\alpha} c_n \\ - \left(\frac{1}{2} \alpha \ddot \alpha + \dot \alpha^2\right) \frac{n_s a^2 n^2}{\lambdabar} c_n \ .
\end{multline*}
Making the cannonical transformation $c'_n = c_n \exp \left(- i e
\int\limits_{0}^{z} \frac{dz'}{\alpha(z')}\right)$ we obtain
Eq. \eqref{discrete}, namely,
\begin{equation*} \label{}
  i \dot c_n = - \frac{k}{\alpha} (c_{n+1} + c_{n-1}) + i \frac{\dot \alpha}{\alpha} c_n - \left(\frac{1}{2} \alpha \ddot \alpha + \dot \alpha^2\right) \frac{n_s a^2 n^2}{\lambdabar} c_n \ .
\end{equation*}

\bibliographystyle{apsrev4-1}
\bibliography{main.bib}

\begin{thebibliography}{33}%
\makeatletter
\providecommand \@ifxundefined [1]{%
 \@ifx{#1\undefined}
}%
\providecommand \@ifnum [1]{%
 \ifnum #1\expandafter \@firstoftwo
 \else \expandafter \@secondoftwo
 \fi
}%
\providecommand \@ifx [1]{%
 \ifx #1\expandafter \@firstoftwo
 \else \expandafter \@secondoftwo
 \fi
}%
\providecommand \natexlab [1]{#1}%
\providecommand \enquote  [1]{``#1''}%
\providecommand \bibnamefont  [1]{#1}%
\providecommand \bibfnamefont [1]{#1}%
\providecommand \citenamefont [1]{#1}%
\providecommand \href@noop [0]{\@secondoftwo}%
\providecommand \href [0]{\begingroup \@sanitize@url \@href}%
\providecommand \@href[1]{\@@startlink{#1}\@@href}%
\providecommand \@@href[1]{\endgroup#1\@@endlink}%
\providecommand \@sanitize@url [0]{\catcode `\\12\catcode `\$12\catcode
  `\&12\catcode `\#12\catcode `\^12\catcode `\_12\catcode `\%12\relax}%
\providecommand \@@startlink[1]{}%
\providecommand \@@endlink[0]{}%
\providecommand \url  [0]{\begingroup\@sanitize@url \@url }%
\providecommand \@url [1]{\endgroup\@href {#1}{\urlprefix }}%
\providecommand \urlprefix  [0]{URL }%
\providecommand \Eprint [0]{\href }%
\providecommand \doibase [0]{http://dx.doi.org/}%
\providecommand \selectlanguage [0]{\@gobble}%
\providecommand \bibinfo  [0]{\@secondoftwo}%
\providecommand \bibfield  [0]{\@secondoftwo}%
\providecommand \translation [1]{[#1]}%
\providecommand \BibitemOpen [0]{}%
\providecommand \bibitemStop [0]{}%
\providecommand \bibitemNoStop [0]{.\EOS\space}%
\providecommand \EOS [0]{\spacefactor3000\relax}%
\providecommand \BibitemShut  [1]{\csname bibitem#1\endcsname}%
\let\auto@bib@innerbib\@empty
%</preamble>
\bibitem [{\citenamefont {Rechtsman}\ \emph {et~al.}(2013)\citenamefont
  {Rechtsman}, \citenamefont {Zeuner}, \citenamefont {Plotnik}, \citenamefont
  {Lumer}, \citenamefont {Podolsky}, \citenamefont {Dreisow}, \citenamefont
  {Nolte}, \citenamefont {Segev},\ and\ \citenamefont
  {Szameit}}]{rechtsman2013photonic}%
  \BibitemOpen
  \bibfield  {author} {\bibinfo {author} {\bibfnamefont {M.~C.}\ \bibnamefont
  {Rechtsman}}, \bibinfo {author} {\bibfnamefont {J.~M.}\ \bibnamefont
  {Zeuner}}, \bibinfo {author} {\bibfnamefont {Y.}~\bibnamefont {Plotnik}},
  \bibinfo {author} {\bibfnamefont {Y.}~\bibnamefont {Lumer}}, \bibinfo
  {author} {\bibfnamefont {D.}~\bibnamefont {Podolsky}}, \bibinfo {author}
  {\bibfnamefont {F.}~\bibnamefont {Dreisow}}, \bibinfo {author} {\bibfnamefont
  {S.}~\bibnamefont {Nolte}}, \bibinfo {author} {\bibfnamefont
  {M.}~\bibnamefont {Segev}}, \ and\ \bibinfo {author} {\bibfnamefont
  {A.}~\bibnamefont {Szameit}},\ }\href@noop {} {\bibfield  {journal} {\bibinfo
   {journal} {Nature}\ }\textbf {\bibinfo {volume} {496}},\ \bibinfo {pages}
  {196} (\bibinfo {year} {2013})}\BibitemShut {NoStop}%
\bibitem [{\citenamefont {Polini}\ \emph {et~al.}(2013)\citenamefont {Polini},
  \citenamefont {Guinea}, \citenamefont {Lewenstein}, \citenamefont
  {Manoharan},\ and\ \citenamefont {Pellegrini}}]{polini2013artificial}%
  \BibitemOpen
  \bibfield  {author} {\bibinfo {author} {\bibfnamefont {M.}~\bibnamefont
  {Polini}}, \bibinfo {author} {\bibfnamefont {F.}~\bibnamefont {Guinea}},
  \bibinfo {author} {\bibfnamefont {M.}~\bibnamefont {Lewenstein}}, \bibinfo
  {author} {\bibfnamefont {H.~C.}\ \bibnamefont {Manoharan}}, \ and\ \bibinfo
  {author} {\bibfnamefont {V.}~\bibnamefont {Pellegrini}},\ }\href@noop {}
  {\bibfield  {journal} {\bibinfo  {journal} {Nature nanotechnology}\ }\textbf
  {\bibinfo {volume} {8}},\ \bibinfo {pages} {625} (\bibinfo {year}
  {2013})}\BibitemShut {NoStop}%
\bibitem [{\citenamefont {Jacqmin}\ \emph {et~al.}(2014)\citenamefont
  {Jacqmin}, \citenamefont {Carusotto}, \citenamefont {Sagnes}, \citenamefont
  {Abbarchi}, \citenamefont {Solnyshkov}, \citenamefont {Malpuech},
  \citenamefont {Galopin}, \citenamefont {Lema{\^\i}tre}, \citenamefont
  {Bloch},\ and\ \citenamefont {Amo}}]{jacqmin2014direct}%
  \BibitemOpen
  \bibfield  {author} {\bibinfo {author} {\bibfnamefont {T.}~\bibnamefont
  {Jacqmin}}, \bibinfo {author} {\bibfnamefont {I.}~\bibnamefont {Carusotto}},
  \bibinfo {author} {\bibfnamefont {I.}~\bibnamefont {Sagnes}}, \bibinfo
  {author} {\bibfnamefont {M.}~\bibnamefont {Abbarchi}}, \bibinfo {author}
  {\bibfnamefont {D.}~\bibnamefont {Solnyshkov}}, \bibinfo {author}
  {\bibfnamefont {G.}~\bibnamefont {Malpuech}}, \bibinfo {author}
  {\bibfnamefont {E.}~\bibnamefont {Galopin}}, \bibinfo {author} {\bibfnamefont
  {A.}~\bibnamefont {Lema{\^\i}tre}}, \bibinfo {author} {\bibfnamefont
  {J.}~\bibnamefont {Bloch}}, \ and\ \bibinfo {author} {\bibfnamefont
  {A.}~\bibnamefont {Amo}},\ }\href@noop {} {\bibfield  {journal} {\bibinfo
  {journal} {Physical review letters}\ }\textbf {\bibinfo {volume} {112}},\
  \bibinfo {pages} {116402} (\bibinfo {year} {2014})}\BibitemShut {NoStop}%
\bibitem [{\citenamefont {Dalibard}\ \emph {et~al.}(2011)\citenamefont
  {Dalibard}, \citenamefont {Gerbier}, \citenamefont {Juzeli{\=u}nas},\ and\
  \citenamefont {{\"O}hberg}}]{dalibard2011colloquium}%
  \BibitemOpen
  \bibfield  {author} {\bibinfo {author} {\bibfnamefont {J.}~\bibnamefont
  {Dalibard}}, \bibinfo {author} {\bibfnamefont {F.}~\bibnamefont {Gerbier}},
  \bibinfo {author} {\bibfnamefont {G.}~\bibnamefont {Juzeli{\=u}nas}}, \ and\
  \bibinfo {author} {\bibfnamefont {P.}~\bibnamefont {{\"O}hberg}},\
  }\href@noop {} {\bibfield  {journal} {\bibinfo  {journal} {Reviews of Modern
  Physics}\ }\textbf {\bibinfo {volume} {83}},\ \bibinfo {pages} {1523}
  (\bibinfo {year} {2011})}\BibitemShut {NoStop}%
\bibitem [{\citenamefont {Zhai}(2012)}]{zhai2012spin}%
  \BibitemOpen
  \bibfield  {author} {\bibinfo {author} {\bibfnamefont {H.}~\bibnamefont
  {Zhai}},\ }\href@noop {} {\bibfield  {journal} {\bibinfo  {journal}
  {International Journal of Modern Physics B}\ }\textbf {\bibinfo {volume}
  {26}},\ \bibinfo {pages} {1230001} (\bibinfo {year} {2012})}\BibitemShut
  {NoStop}%
\bibitem [{\citenamefont {Longhi}\ \emph {et~al.}(2006)\citenamefont {Longhi},
  \citenamefont {Marangoni}, \citenamefont {Lobino}, \citenamefont {Ramponi},
  \citenamefont {Laporta}, \citenamefont {Cianci},\ and\ \citenamefont
  {Foglietti}}]{longhi2006observation}%
  \BibitemOpen
  \bibfield  {author} {\bibinfo {author} {\bibfnamefont {S.}~\bibnamefont
  {Longhi}}, \bibinfo {author} {\bibfnamefont {M.}~\bibnamefont {Marangoni}},
  \bibinfo {author} {\bibfnamefont {M.}~\bibnamefont {Lobino}}, \bibinfo
  {author} {\bibfnamefont {R.}~\bibnamefont {Ramponi}}, \bibinfo {author}
  {\bibfnamefont {P.}~\bibnamefont {Laporta}}, \bibinfo {author} {\bibfnamefont
  {E.}~\bibnamefont {Cianci}}, \ and\ \bibinfo {author} {\bibfnamefont
  {V.}~\bibnamefont {Foglietti}},\ }\href@noop {} {\bibfield  {journal}
  {\bibinfo  {journal} {Physical review letters}\ }\textbf {\bibinfo {volume}
  {96}},\ \bibinfo {pages} {243901} (\bibinfo {year} {2006})}\BibitemShut
  {NoStop}%
\bibitem [{\citenamefont {Szameit}\ \emph {et~al.}(2010)\citenamefont
  {Szameit}, \citenamefont {Garanovich}, \citenamefont {Heinrich},
  \citenamefont {Sukhorukov}, \citenamefont {Dreisow}, \citenamefont {Pertsch},
  \citenamefont {Nolte}, \citenamefont {T{\"u}nnermann}, \citenamefont
  {Longhi},\ and\ \citenamefont {Kivshar}}]{szameit2010observation}%
  \BibitemOpen
  \bibfield  {author} {\bibinfo {author} {\bibfnamefont {A.}~\bibnamefont
  {Szameit}}, \bibinfo {author} {\bibfnamefont {I.~L.}\ \bibnamefont
  {Garanovich}}, \bibinfo {author} {\bibfnamefont {M.}~\bibnamefont
  {Heinrich}}, \bibinfo {author} {\bibfnamefont {A.~A.}\ \bibnamefont
  {Sukhorukov}}, \bibinfo {author} {\bibfnamefont {F.}~\bibnamefont {Dreisow}},
  \bibinfo {author} {\bibfnamefont {T.}~\bibnamefont {Pertsch}}, \bibinfo
  {author} {\bibfnamefont {S.}~\bibnamefont {Nolte}}, \bibinfo {author}
  {\bibfnamefont {A.}~\bibnamefont {T{\"u}nnermann}}, \bibinfo {author}
  {\bibfnamefont {S.}~\bibnamefont {Longhi}}, \ and\ \bibinfo {author}
  {\bibfnamefont {Y.~S.}\ \bibnamefont {Kivshar}},\ }\href@noop {} {\bibfield
  {journal} {\bibinfo  {journal} {Physical review letters}\ }\textbf {\bibinfo
  {volume} {104}},\ \bibinfo {pages} {223903} (\bibinfo {year}
  {2010})}\BibitemShut {NoStop}%
\bibitem [{\citenamefont {Longhi}(2015)}]{longhi2015localization}%
  \BibitemOpen
  \bibfield  {author} {\bibinfo {author} {\bibfnamefont {S.}~\bibnamefont
  {Longhi}},\ }\href@noop {} {\bibfield  {journal} {\bibinfo  {journal} {Optics
  letters}\ }\textbf {\bibinfo {volume} {40}},\ \bibinfo {pages} {4707}
  (\bibinfo {year} {2015})}\BibitemShut {NoStop}%
\bibitem [{\citenamefont {Longhi}(2005{\natexlab{a}})}]{longhi2005coherent}%
  \BibitemOpen
  \bibfield  {author} {\bibinfo {author} {\bibfnamefont {S.}~\bibnamefont
  {Longhi}},\ }\href@noop {} {\bibfield  {journal} {\bibinfo  {journal}
  {Physical Review A}\ }\textbf {\bibinfo {volume} {71}},\ \bibinfo {pages}
  {065801} (\bibinfo {year} {2005}{\natexlab{a}})}\BibitemShut {NoStop}%
\bibitem [{\citenamefont {Della~Valle}\ \emph {et~al.}(2007)\citenamefont
  {Della~Valle}, \citenamefont {Ornigotti}, \citenamefont {Cianci},
  \citenamefont {Foglietti}, \citenamefont {Laporta},\ and\ \citenamefont
  {Longhi}}]{della2007visualization}%
  \BibitemOpen
  \bibfield  {author} {\bibinfo {author} {\bibfnamefont {G.}~\bibnamefont
  {Della~Valle}}, \bibinfo {author} {\bibfnamefont {M.}~\bibnamefont
  {Ornigotti}}, \bibinfo {author} {\bibfnamefont {E.}~\bibnamefont {Cianci}},
  \bibinfo {author} {\bibfnamefont {V.}~\bibnamefont {Foglietti}}, \bibinfo
  {author} {\bibfnamefont {P.}~\bibnamefont {Laporta}}, \ and\ \bibinfo
  {author} {\bibfnamefont {S.}~\bibnamefont {Longhi}},\ }\href@noop {}
  {\bibfield  {journal} {\bibinfo  {journal} {Physical review letters}\
  }\textbf {\bibinfo {volume} {98}},\ \bibinfo {pages} {263601} (\bibinfo
  {year} {2007})}\BibitemShut {NoStop}%
\bibitem [{\citenamefont {Szameit}\ \emph {et~al.}(2009)\citenamefont
  {Szameit}, \citenamefont {Kartashov}, \citenamefont {Dreisow}, \citenamefont
  {Heinrich}, \citenamefont {Pertsch}, \citenamefont {Nolte}, \citenamefont
  {T{\"u}nnermann}, \citenamefont {Vysloukh}, \citenamefont {Lederer},\ and\
  \citenamefont {Torner}}]{szameit2009inhibition}%
  \BibitemOpen
  \bibfield  {author} {\bibinfo {author} {\bibfnamefont {A.}~\bibnamefont
  {Szameit}}, \bibinfo {author} {\bibfnamefont {Y.~V.}\ \bibnamefont
  {Kartashov}}, \bibinfo {author} {\bibfnamefont {F.}~\bibnamefont {Dreisow}},
  \bibinfo {author} {\bibfnamefont {M.}~\bibnamefont {Heinrich}}, \bibinfo
  {author} {\bibfnamefont {T.}~\bibnamefont {Pertsch}}, \bibinfo {author}
  {\bibfnamefont {S.}~\bibnamefont {Nolte}}, \bibinfo {author} {\bibfnamefont
  {A.}~\bibnamefont {T{\"u}nnermann}}, \bibinfo {author} {\bibfnamefont
  {V.~A.}\ \bibnamefont {Vysloukh}}, \bibinfo {author} {\bibfnamefont
  {F.}~\bibnamefont {Lederer}}, \ and\ \bibinfo {author} {\bibfnamefont
  {L.}~\bibnamefont {Torner}},\ }\href@noop {} {\bibfield  {journal} {\bibinfo
  {journal} {Physical review letters}\ }\textbf {\bibinfo {volume} {102}},\
  \bibinfo {pages} {153901} (\bibinfo {year} {2009})}\BibitemShut {NoStop}%
\bibitem [{\citenamefont {Mukherjee}\ \emph {et~al.}(2016)\citenamefont
  {Mukherjee}, \citenamefont {Valiente}, \citenamefont {Goldman}, \citenamefont
  {Spracklen}, \citenamefont {Andersson}, \citenamefont {{\"O}hberg},\ and\
  \citenamefont {Thomson}}]{mukherjee2016observation}%
  \BibitemOpen
  \bibfield  {author} {\bibinfo {author} {\bibfnamefont {S.}~\bibnamefont
  {Mukherjee}}, \bibinfo {author} {\bibfnamefont {M.}~\bibnamefont {Valiente}},
  \bibinfo {author} {\bibfnamefont {N.}~\bibnamefont {Goldman}}, \bibinfo
  {author} {\bibfnamefont {A.}~\bibnamefont {Spracklen}}, \bibinfo {author}
  {\bibfnamefont {E.}~\bibnamefont {Andersson}}, \bibinfo {author}
  {\bibfnamefont {P.}~\bibnamefont {{\"O}hberg}}, \ and\ \bibinfo {author}
  {\bibfnamefont {R.~R.}\ \bibnamefont {Thomson}},\ }\href@noop {} {\bibfield
  {journal} {\bibinfo  {journal} {Physical Review A}\ }\textbf {\bibinfo
  {volume} {94}},\ \bibinfo {pages} {053853} (\bibinfo {year}
  {2016})}\BibitemShut {NoStop}%
\bibitem [{\citenamefont {Luo}\ \emph {et~al.}(2007)\citenamefont {Luo},
  \citenamefont {Xie},\ and\ \citenamefont {Wu}}]{luo2007nonlinear}%
  \BibitemOpen
  \bibfield  {author} {\bibinfo {author} {\bibfnamefont {X.}~\bibnamefont
  {Luo}}, \bibinfo {author} {\bibfnamefont {Q.}~\bibnamefont {Xie}}, \ and\
  \bibinfo {author} {\bibfnamefont {B.}~\bibnamefont {Wu}},\ }\href@noop {}
  {\bibfield  {journal} {\bibinfo  {journal} {Physical Review A}\ }\textbf
  {\bibinfo {volume} {76}},\ \bibinfo {pages} {051802} (\bibinfo {year}
  {2007})}\BibitemShut {NoStop}%
\bibitem [{\citenamefont {Ornigotti}\ \emph {et~al.}(2008)\citenamefont
  {Ornigotti}, \citenamefont {Della~Valle}, \citenamefont {Fernandez},
  \citenamefont {Coppa}, \citenamefont {Foglietti}, \citenamefont {Laporta},\
  and\ \citenamefont {Longhi}}]{ornigotti2008visualization}%
  \BibitemOpen
  \bibfield  {author} {\bibinfo {author} {\bibfnamefont {M.}~\bibnamefont
  {Ornigotti}}, \bibinfo {author} {\bibfnamefont {G.}~\bibnamefont
  {Della~Valle}}, \bibinfo {author} {\bibfnamefont {T.~T.}\ \bibnamefont
  {Fernandez}}, \bibinfo {author} {\bibfnamefont {A.}~\bibnamefont {Coppa}},
  \bibinfo {author} {\bibfnamefont {V.}~\bibnamefont {Foglietti}}, \bibinfo
  {author} {\bibfnamefont {P.}~\bibnamefont {Laporta}}, \ and\ \bibinfo
  {author} {\bibfnamefont {S.}~\bibnamefont {Longhi}},\ }\href@noop {}
  {\bibfield  {journal} {\bibinfo  {journal} {Journal of Physics B: Atomic,
  Molecular and Optical Physics}\ }\textbf {\bibinfo {volume} {41}},\ \bibinfo
  {pages} {085402} (\bibinfo {year} {2008})}\BibitemShut {NoStop}%
\bibitem [{\citenamefont {Kartashov}\ \emph {et~al.}(2007)\citenamefont
  {Kartashov}, \citenamefont {Vysloukh},\ and\ \citenamefont
  {Torner}}]{PhysRevLett.99.233903}%
  \BibitemOpen
  \bibfield  {author} {\bibinfo {author} {\bibfnamefont {Y.~V.}\ \bibnamefont
  {Kartashov}}, \bibinfo {author} {\bibfnamefont {V.~A.}\ \bibnamefont
  {Vysloukh}}, \ and\ \bibinfo {author} {\bibfnamefont {L.}~\bibnamefont
  {Torner}},\ }\href {\doibase 10.1103/PhysRevLett.99.233903} {\bibfield
  {journal} {\bibinfo  {journal} {Phys. Rev. Lett.}\ }\textbf {\bibinfo
  {volume} {99}},\ \bibinfo {pages} {233903} (\bibinfo {year}
  {2007})}\BibitemShut {NoStop}%
\bibitem [{\citenamefont {Shandarova}\ \emph {et~al.}(2009)\citenamefont
  {Shandarova}, \citenamefont {R\"uter}, \citenamefont {Kip}, \citenamefont
  {Makris}, \citenamefont {Christodoulides}, \citenamefont {Peleg},\ and\
  \citenamefont {Segev}}]{PhysRevLett.102.123905}%
  \BibitemOpen
  \bibfield  {author} {\bibinfo {author} {\bibfnamefont {K.}~\bibnamefont
  {Shandarova}}, \bibinfo {author} {\bibfnamefont {C.~E.}\ \bibnamefont
  {R\"uter}}, \bibinfo {author} {\bibfnamefont {D.}~\bibnamefont {Kip}},
  \bibinfo {author} {\bibfnamefont {K.~G.}\ \bibnamefont {Makris}}, \bibinfo
  {author} {\bibfnamefont {D.~N.}\ \bibnamefont {Christodoulides}}, \bibinfo
  {author} {\bibfnamefont {O.}~\bibnamefont {Peleg}}, \ and\ \bibinfo {author}
  {\bibfnamefont {M.}~\bibnamefont {Segev}},\ }\href {\doibase
  10.1103/PhysRevLett.102.123905} {\bibfield  {journal} {\bibinfo  {journal}
  {Phys. Rev. Lett.}\ }\textbf {\bibinfo {volume} {102}},\ \bibinfo {pages}
  {123905} (\bibinfo {year} {2009})}\BibitemShut {NoStop}%
\bibitem [{\citenamefont {Schwartz}(2007)}]{schwartz2007t}%
  \BibitemOpen
  \bibfield  {author} {\bibinfo {author} {\bibfnamefont {T.}~\bibnamefont
  {Schwartz}},\ }\href@noop {} {\bibfield  {journal} {\bibinfo  {journal}
  {Nature (London)}\ }\textbf {\bibinfo {volume} {446}},\ \bibinfo {pages} {52}
  (\bibinfo {year} {2007})}\BibitemShut {NoStop}%
\bibitem [{\citenamefont {Wiersma}\ \emph {et~al.}(1997)\citenamefont
  {Wiersma}, \citenamefont {Bartolini}, \citenamefont {Lagendijk},\ and\
  \citenamefont {Righini}}]{wiersma1997localization}%
  \BibitemOpen
  \bibfield  {author} {\bibinfo {author} {\bibfnamefont {D.~S.}\ \bibnamefont
  {Wiersma}}, \bibinfo {author} {\bibfnamefont {P.}~\bibnamefont {Bartolini}},
  \bibinfo {author} {\bibfnamefont {A.}~\bibnamefont {Lagendijk}}, \ and\
  \bibinfo {author} {\bibfnamefont {R.}~\bibnamefont {Righini}},\ }\href@noop
  {} {\bibfield  {journal} {\bibinfo  {journal} {Nature}\ }\textbf {\bibinfo
  {volume} {390}},\ \bibinfo {pages} {671} (\bibinfo {year}
  {1997})}\BibitemShut {NoStop}%
\bibitem [{\citenamefont {Lahini}\ \emph {et~al.}(2008)\citenamefont {Lahini},
  \citenamefont {Avidan}, \citenamefont {Pozzi}, \citenamefont {Sorel},
  \citenamefont {Morandotti}, \citenamefont {Christodoulides},\ and\
  \citenamefont {Silberberg}}]{lahini2008anderson}%
  \BibitemOpen
  \bibfield  {author} {\bibinfo {author} {\bibfnamefont {Y.}~\bibnamefont
  {Lahini}}, \bibinfo {author} {\bibfnamefont {A.}~\bibnamefont {Avidan}},
  \bibinfo {author} {\bibfnamefont {F.}~\bibnamefont {Pozzi}}, \bibinfo
  {author} {\bibfnamefont {M.}~\bibnamefont {Sorel}}, \bibinfo {author}
  {\bibfnamefont {R.}~\bibnamefont {Morandotti}}, \bibinfo {author}
  {\bibfnamefont {D.~N.}\ \bibnamefont {Christodoulides}}, \ and\ \bibinfo
  {author} {\bibfnamefont {Y.}~\bibnamefont {Silberberg}},\ }\href@noop {}
  {\bibfield  {journal} {\bibinfo  {journal} {Physical Review Letters}\
  }\textbf {\bibinfo {volume} {100}},\ \bibinfo {pages} {013906} (\bibinfo
  {year} {2008})}\BibitemShut {NoStop}%
\bibitem [{\citenamefont {Wolf}\ and\ \citenamefont
  {Maret}(1985)}]{wolf1985weak}%
  \BibitemOpen
  \bibfield  {author} {\bibinfo {author} {\bibfnamefont {P.-E.}\ \bibnamefont
  {Wolf}}\ and\ \bibinfo {author} {\bibfnamefont {G.}~\bibnamefont {Maret}},\
  }\href@noop {} {\bibfield  {journal} {\bibinfo  {journal} {Physical review
  letters}\ }\textbf {\bibinfo {volume} {55}},\ \bibinfo {pages} {2696}
  (\bibinfo {year} {1985})}\BibitemShut {NoStop}%
\bibitem [{\citenamefont {Van~Albada}\ and\ \citenamefont
  {Lagendijk}(1985)}]{van1985observation}%
  \BibitemOpen
  \bibfield  {author} {\bibinfo {author} {\bibfnamefont {M.~P.}\ \bibnamefont
  {Van~Albada}}\ and\ \bibinfo {author} {\bibfnamefont {A.}~\bibnamefont
  {Lagendijk}},\ }\href@noop {} {\bibfield  {journal} {\bibinfo  {journal}
  {Physical review letters}\ }\textbf {\bibinfo {volume} {55}},\ \bibinfo
  {pages} {2692} (\bibinfo {year} {1985})}\BibitemShut {NoStop}%
\bibitem [{\citenamefont {John}(1984)}]{john1984electromagnetic}%
  \BibitemOpen
  \bibfield  {author} {\bibinfo {author} {\bibfnamefont {S.}~\bibnamefont
  {John}},\ }\href@noop {} {\bibfield  {journal} {\bibinfo  {journal} {Physical
  Review Letters}\ }\textbf {\bibinfo {volume} {53}},\ \bibinfo {pages} {2169}
  (\bibinfo {year} {1984})}\BibitemShut {NoStop}%
\bibitem [{\citenamefont {Theocharis}\ \emph {et~al.}(2006)\citenamefont
  {Theocharis}, \citenamefont {Schmelcher}, \citenamefont {Kevrekidis},\ and\
  \citenamefont {Frantzeskakis}}]{theocharis2006dynamical}%
  \BibitemOpen
  \bibfield  {author} {\bibinfo {author} {\bibfnamefont {G.}~\bibnamefont
  {Theocharis}}, \bibinfo {author} {\bibfnamefont {P.}~\bibnamefont
  {Schmelcher}}, \bibinfo {author} {\bibfnamefont {P.}~\bibnamefont
  {Kevrekidis}}, \ and\ \bibinfo {author} {\bibfnamefont {D.}~\bibnamefont
  {Frantzeskakis}},\ }\href@noop {} {\bibfield  {journal} {\bibinfo  {journal}
  {Physical Review A}\ }\textbf {\bibinfo {volume} {74}},\ \bibinfo {pages}
  {053614} (\bibinfo {year} {2006})}\BibitemShut {NoStop}%
\bibitem [{\citenamefont {Longhi}(2011)}]{longhi2011dynamic}%
  \BibitemOpen
  \bibfield  {author} {\bibinfo {author} {\bibfnamefont {S.}~\bibnamefont
  {Longhi}},\ }\href@noop {} {\bibfield  {journal} {\bibinfo  {journal} {Optics
  letters}\ }\textbf {\bibinfo {volume} {36}},\ \bibinfo {pages} {819}
  (\bibinfo {year} {2011})}\BibitemShut {NoStop}%
\bibitem [{\citenamefont {Papagiannis}\ \emph {et~al.}(2011)\citenamefont
  {Papagiannis}, \citenamefont {Kominis},\ and\ \citenamefont
  {Hizanidis}}]{papagiannis2011power}%
  \BibitemOpen
  \bibfield  {author} {\bibinfo {author} {\bibfnamefont {P.}~\bibnamefont
  {Papagiannis}}, \bibinfo {author} {\bibfnamefont {Y.}~\bibnamefont
  {Kominis}}, \ and\ \bibinfo {author} {\bibfnamefont {K.}~\bibnamefont
  {Hizanidis}},\ }\href@noop {} {\bibfield  {journal} {\bibinfo  {journal}
  {Physical Review A}\ }\textbf {\bibinfo {volume} {84}},\ \bibinfo {pages}
  {013820} (\bibinfo {year} {2011})}\BibitemShut {NoStop}%
\bibitem [{\citenamefont {Longhi}(2005{\natexlab{b}})}]{longhi2005self}%
  \BibitemOpen
  \bibfield  {author} {\bibinfo {author} {\bibfnamefont {S.}~\bibnamefont
  {Longhi}},\ }\href@noop {} {\bibfield  {journal} {\bibinfo  {journal} {Optics
  letters}\ }\textbf {\bibinfo {volume} {30}},\ \bibinfo {pages} {2137}
  (\bibinfo {year} {2005}{\natexlab{b}})}\BibitemShut {NoStop}%
\bibitem [{\citenamefont {Iyer}\ \emph {et~al.}(2007)\citenamefont {Iyer},
  \citenamefont {Aitchison}, \citenamefont {Wan}, \citenamefont {Dignam},\ and\
  \citenamefont {de~Sterke}}]{iyer2007exact}%
  \BibitemOpen
  \bibfield  {author} {\bibinfo {author} {\bibfnamefont {R.}~\bibnamefont
  {Iyer}}, \bibinfo {author} {\bibfnamefont {J.~S.}\ \bibnamefont {Aitchison}},
  \bibinfo {author} {\bibfnamefont {J.}~\bibnamefont {Wan}}, \bibinfo {author}
  {\bibfnamefont {M.~M.}\ \bibnamefont {Dignam}}, \ and\ \bibinfo {author}
  {\bibfnamefont {C.~M.}\ \bibnamefont {de~Sterke}},\ }\href@noop {} {\bibfield
   {journal} {\bibinfo  {journal} {Optics express}\ }\textbf {\bibinfo {volume}
  {15}},\ \bibinfo {pages} {3212} (\bibinfo {year} {2007})}\BibitemShut
  {NoStop}%
\bibitem [{\citenamefont {Earnshaw}(1842)}]{earnshaw1842nature}%
  \BibitemOpen
  \bibfield  {author} {\bibinfo {author} {\bibfnamefont {S.}~\bibnamefont
  {Earnshaw}},\ }\href@noop {} {\bibfield  {journal} {\bibinfo  {journal}
  {Trans. Camb. Phil. Soc.}\ }\textbf {\bibinfo {volume} {7}},\ \bibinfo
  {pages} {97} (\bibinfo {year} {1842})}\BibitemShut {NoStop}%
\bibitem [{\citenamefont {Goldman}\ and\ \citenamefont
  {Dalibard}(2014)}]{goldman2014periodically}%
  \BibitemOpen
  \bibfield  {author} {\bibinfo {author} {\bibfnamefont {N.}~\bibnamefont
  {Goldman}}\ and\ \bibinfo {author} {\bibfnamefont {J.}~\bibnamefont
  {Dalibard}},\ }\href@noop {} {\bibfield  {journal} {\bibinfo  {journal}
  {Physical review X}\ }\textbf {\bibinfo {volume} {4}},\ \bibinfo {pages}
  {031027} (\bibinfo {year} {2014})}\BibitemShut {NoStop}%
\bibitem [{\citenamefont {Cook}\ \emph {et~al.}(1985)\citenamefont {Cook},
  \citenamefont {Shankland},\ and\ \citenamefont {Wells}}]{cook1985quantum}%
  \BibitemOpen
  \bibfield  {author} {\bibinfo {author} {\bibfnamefont {R.~J.}\ \bibnamefont
  {Cook}}, \bibinfo {author} {\bibfnamefont {D.~G.}\ \bibnamefont {Shankland}},
  \ and\ \bibinfo {author} {\bibfnamefont {A.~L.}\ \bibnamefont {Wells}},\
  }\href@noop {} {\bibfield  {journal} {\bibinfo  {journal} {Physical Review
  A}\ }\textbf {\bibinfo {volume} {31}},\ \bibinfo {pages} {564} (\bibinfo
  {year} {1985})}\BibitemShut {NoStop}%
\bibitem [{\citenamefont {Gilary}\ \emph {et~al.}(2003)\citenamefont {Gilary},
  \citenamefont {Moiseyev}, \citenamefont {Rahav},\ and\ \citenamefont
  {Fishman}}]{gilary2003trapping}%
  \BibitemOpen
  \bibfield  {author} {\bibinfo {author} {\bibfnamefont {I.}~\bibnamefont
  {Gilary}}, \bibinfo {author} {\bibfnamefont {N.}~\bibnamefont {Moiseyev}},
  \bibinfo {author} {\bibfnamefont {S.}~\bibnamefont {Rahav}}, \ and\ \bibinfo
  {author} {\bibfnamefont {S.}~\bibnamefont {Fishman}},\ }\href@noop {}
  {\bibfield  {journal} {\bibinfo  {journal} {Journal of Physics A:
  Mathematical and General}\ }\textbf {\bibinfo {volume} {36}},\ \bibinfo
  {pages} {L409} (\bibinfo {year} {2003})}\BibitemShut {NoStop}%
\bibitem [{\citenamefont {Carrasco}\ \emph {et~al.}(2017)\citenamefont
  {Carrasco}, \citenamefont {Rogan},\ and\ \citenamefont
  {Valdivia}}]{carrasco2017}%
  \BibitemOpen
  \bibfield  {author} {\bibinfo {author} {\bibfnamefont {S.}~\bibnamefont
  {Carrasco}}, \bibinfo {author} {\bibfnamefont {J.}~\bibnamefont {Rogan}}, \
  and\ \bibinfo {author} {\bibfnamefont {J.~A.}\ \bibnamefont {Valdivia}},\
  }\href@noop {} {\bibfield  {journal} {\bibinfo  {journal} {Scientific
  reports}\ }\textbf {\bibinfo {volume} {7}},\ \bibinfo {pages} {13217}
  (\bibinfo {year} {2017})}\BibitemShut {NoStop}%
\bibitem [{\citenamefont {Weimann}\ \emph {et~al.}(2018)\citenamefont
  {Weimann}, \citenamefont {Eichelkraut},\ and\ \citenamefont
  {Szameit}}]{weimann2018decay}%
  \BibitemOpen
  \bibfield  {author} {\bibinfo {author} {\bibfnamefont {S.}~\bibnamefont
  {Weimann}}, \bibinfo {author} {\bibfnamefont {T.}~\bibnamefont
  {Eichelkraut}}, \ and\ \bibinfo {author} {\bibfnamefont {A.}~\bibnamefont
  {Szameit}},\ }\href@noop {} {\bibfield  {journal} {\bibinfo  {journal}
  {Physical Review A}\ }\textbf {\bibinfo {volume} {97}},\ \bibinfo {pages}
  {053844} (\bibinfo {year} {2018})}\BibitemShut {NoStop}%
\end{thebibliography}%

\end{document}